\documentclass[
  ,final            
  ]
  {aipproc}

\usepackage{amsmath}

\layoutstyle{8x11double}

\begin{document}

\title{Continuum driven winds from super-Eddington stars. A tale of two limits}


\author{A.~J. van Marle}{
  address={Bartol Research Institute, University of Delaware, Newark, DE  19716, USA}
}

\author{S.~P. Owocki}{
  address={Bartol Research Institute, University of Delaware, Newark, DE  19716, USA}
}

\author{N.~J. Shaviv}{
  address={Racah Institute of Physics, Hebrew University, Giv'at Ram, Jerusalem 91904, Israel}
}

\begin{abstract}
Continuum driving is an effective method to drive a strong stellar wind. It is governed by two limits: the 
Eddington limit and the photon-tiring limit. A star must exceed the effective Eddington limit for continuum driving to 
overcome the stellar gravity. The photon-tiring limit places an upper limit on the mass loss rate that can be driven to 
infinity, given the energy available in the radiation field of the star. 
 
Since continuum driving does not require the presence of metals in the stellar atmosphere it is particularly suited 
to removing mass from low- and zero-metallicity stars and can play a crucial part in their evolution.

Using a porosity length formalism we compute numerical simulations of super-Eddington, continuum driven winds to 
explore their behaviour for 
stars both below and above the photon-tiring limit. We find that below the photon tiring limit, continuum driving can 
produce a large, steady mass loss rate at velocities on the order of the escape velocity. If the star exceeds the 
photon-tiring limit, a steady solution is no longer possible. While the effective mass loss rate is still very large, 
the wind velocity is much smaller.

\end{abstract}

\maketitle


\section{Introduction}
Continuum driven stellar winds are comparatively rare. The star has to exceed the effective Eddington limit for 
continuum driving to be effective at all, with the effective Eddington limit defined as the point where gravity is balanced by the continuum force including both electron scattering and other continuum processes (i.e. bound-free and free-free). Since most stars stay below this limit their winds are driven by line absorption of 
radiation rather than continuum scattering. 

However, continuum driving has two characteristics that may yet make it of great importance for the evolution of zero- 
and low metallicity stars: 1) Unlike line driving, continuum driving does not depend on the presence of metals. Rather 
it depends on the presence of free electrons as well as bound-free interactions, both of which can take place in any 
gas, irrespective of its composition. 2) Continuum driving can propel a potentially far greater mass loss rate than 
line driving. Whereas line driving is limited to about $10^{-4}-10^{-3}~M_\odot~yr^{-1}$ because of line saturation, 
continuum driving can easily drive a hundred times more mass. In fact, the only limit to the mass loss rate of a 
continuum driven wind is the total amount of energy that is available in the radiation field of the star. For line 
driven winds, this upper limit is irrelevant, since the mechanical luminosity of such winds is always much less than 
the radiative luminosity of the star. The high mass loss rate makes continuum driving ideal for explaining massive 
outbursts like the one $\eta$~Carinae experienced between 1840-1860 
(Davidson \& Humphreys \cite{DH97}; Smith \cite{S02}).

The main problem in producing a steady, continuum driven wind lies in the fact that the ratio of radiative flux and 
gravitational acceleration tends to be constant throughout the star. (Both scale with $r^{-2}$.) Therefore, if the star 
is super-Eddington at the surface, it will be super-Eddington throughout its envelope, which leads to large scale 
instabilities as the envelope becomes gravitationally unbound. 

A potential solution to this problem is provided by the "porosity" of the material. This decouples radiation from 
matter as 
photons escape through the low density medium, reducing the interaction with the main body of the gas, which is tied up 
in 
optically thick clumps (Shaviv \cite{Sh98}, \cite{Sh00}).

Owocki et al. \cite{OGS04}, henceforth OGS, worked out analytically what a porosity modified, continuum driven wind 
would look like as 
long as the star does not exceed the photon-tiring limit. Here we present numerical models of similar winds, also 
exploring what happens when a star does in fact exceed the photon tiring limit, which can not be described 
analytically.

\section{Analytical and numerical background}
For our simulations we assume that the stellar atmosphere consists of a porous material, where optically thick clumps 
are surrounded by a low density, optically thin gas. This will cause the photons to escape through the optically thin 
medium, moderating the interaction with the gas in the clumps. As a result, the coupling between radiation and matter 
is 
reduced. This effect decreases as the clumps themselves become optically thin, making the interaction between 
radiation and matter density-dependent. As a result, the full force of the radiation field only comes to bear on the 
gas in the low density outer layers of the stellar atmosphere.

\subsection{Analytical solution}
The acceleration  produced by radiation force on matter through continuum interaction is given by: 
\begin{equation}
g_{\rm rad}~=~ \kappa \frac{F}{c},
\end{equation}
with $\kappa$ the continuum opacity, $F$ the radiative flux and $c$ the lightspeed. If the medium is porous, the 
standard 
opacity has to be replaced by a new, effective opacity, which scales with the optical depth of the individual clumps. 
This in turn depends on the density of the medium. Rather than giving all clumps the same optical depth, we assume a 
power-law distribution, which leads to:
\begin{equation}
\begin{aligned}
g_{\rm rad}~&\rightarrow~\frac{\kappa_{\rm eff}}{\kappa} g_{\rm rad}, \\
            &=\frac{(1+\tau_0)^{1-\alpha_p} - 1}{(1-\alpha_p)\tau_0} g_{\rm rad},
\label{eq:grad}
\end{aligned}
\end{equation}
with $\alpha_p$ the power-law index and $\tau_0\equiv\rho/\rho_0$, with $\rho_0=1/(h \kappa)$ the density at which the largest clump has optical depth unity. This depends on the opacity and the porosity length: $h$ (OGS), which is related to the size of the clumps relative to the distance between clumps.

Since the total energy must be conserved, the mechanical luminosity of the wind may not exceed the radiative luminosity 
of the star. The maximum mass loss rate that can be driven to infinity is therefore given by
\begin{equation}
\begin{aligned}
\dot{M}_{\rm tir}~&=~\frac{2 L_\star}{v_{\rm esc}^2} \\
                  &=~3.18\times 10^{-8}                
\biggl(\frac{L_\star}{L_\odot}\biggr)\biggl(\frac{R_\star}{R_\odot}\biggr)\biggl(\frac{M_\star}{M_\odot}\biggr)^{-1}
                  ~\biggl[\frac{M_\odot} {\rm yr}\biggr],
\end{aligned}
\end{equation}
where $\dot{M}_{\rm tir}$ is the photon-tiring limit of the mass loss rate. If the mass loss rate at the 
stellar surface exceeds this limit, part of the mass will fall back toward the star. For a more thorough explanation of 
the analytical background of the porosity length formalism, as well as the continuum driving force, see OGS.

\subsection{Numerical approach}
We simulate the continuum driving by adding an acceleration term with the scaling of Eq.~\ref{eq:grad} to the momentum
equation. The effect of photon tiring is included by reducing the local radiative luminosity in each gridcell by the 
amount of 
work required to drive the matter to that distance:
\begin{equation}
L(r)~=~L_\star - \int_{R_\star}^r \dot{M}(r') g_{\rm rad}(r') dr'
\label{eq:tir}
\end{equation}
with $L(r)$ the available local luminosity and ${\dot{M}(r')=4\pi r'^2\rho(r')v(r')}$. For a steady wind solution, 
$\dot{M}(r')$ is a constant value, but if the star exceeds the photon-tiring limit this will no longer be the case.
Note that Eq.~\ref{eq:tir} works two ways. If the local velocity is positive, the luminosity is reduced. If the 
velocity is negative, the integral term  becomes negative as well and energy is added to the radiation field. 
In this equation we ignore for simplicity that the amount of work done by gas pressure is small compared to what is 
done by radiative driving.

\section{Numerical result}
For our sample simulations we use a 50$M_\odot$, 50$R_\odot$ star with a surface temperature of 50\,000~K. We ran 
several simulations for different luminosities to explore the behaviour of continuum driven stellar winds.

\begin{figure}
  \includegraphics[width=\columnwidth]{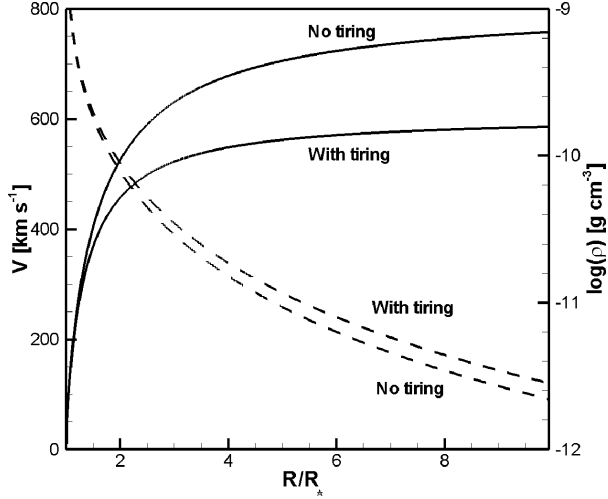}
  \caption{Wind velocity (solid lines) for a $\Gamma=3$, $\alpha=0.5$ simulation of a continuum driven wind from a 
50$M_\odot$, 50$R_\odot$ star. The photon tiring effect only makes a small difference in the terminal velocity. }
\label{fig:subtir1}
\end{figure}

\begin{figure}
  \includegraphics[width=\columnwidth]{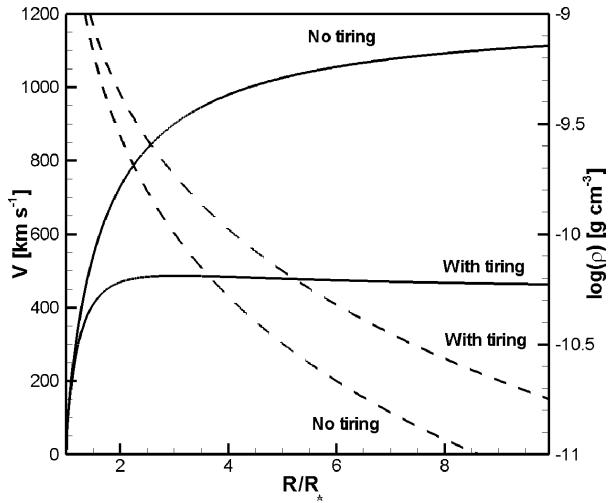}
  \caption{Similar to Fig.~\ref{fig:subtir1}, but with $\Gamma=6$. For this luminosity the photon tiring effect changes the wind velocity considerably. Note that the velocity actually decreases somewhat in the outer layers, as the local 
radiation force becomes less than gravity.}
\label{fig:subtir2}
\end{figure}

\subsection{Below the photon tiring limit}
Figures~\ref{fig:subtir1} and \ref{fig:subtir2} show the radial velocity and density of the wind for 
${\Gamma=L_\star/L_{\rm Edd}=3}$ and 6 respectively, both with and without the effect of photon-tiring included. As can 
be seen in these plots, the difference is comparatively small for the lower luminosity, reducing the wind velocity by 
about 25\%. However, for the high luminosity, the change is considerable. Here the terminal velocity of the wind is 
reduced by more than 50\% and the wind velocity actually decreases somewhat in the outer layers as the local radiation 
field becomes too `tired' to counter balance the force of gravity.
In both cases the terminal velocity of the wind is on the order of the escape velocity at the stellar surface.

The mass loss rates for both winds is high. For ${\Gamma=3}$ the mass loss rate is about $0.04~M_\odot~yr^{-1}$. For 
${\Gamma=6}$ it is approximately $0.2~M_\odot~yr^{-1}$. Such mass loss rates are several orders of magnitude more than 
can be achieved through line driving. Note that photon tiring does not change the mass loss rate. This depends 
exclusively on the porosity parameters and luminosity at the sonic point (OGS), which is defined as the point where $g_{\rm rad}=g_{\rm grav}$. Only if this mass loss rate exceeds 
the photon-tiring limit does the mass loss rate driven to infinity depend on whether the photon tiring effect is 
included.

For these simulations the porosity length was set to the hydrodynamical scale-height at the stellar surface.

\begin{figure}
  \includegraphics[width=\columnwidth]{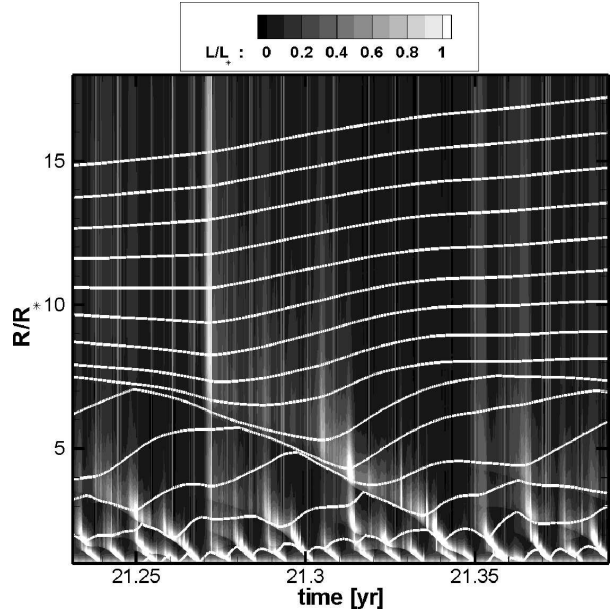}
  \caption{Similar stellar mass and radius as in previous figures, but luminosity at $\Gamma=10$ in order to produce a 
surface mass loss rate of ${\dot{M}=2.34\dot{M}_{\rm tir}}$. The line contours show the motion of individual mass 
elements, while the surface contour in the background shows the local radiative luminosity; both as functions of time 
and radius. The mass elements that leave the surface fall back down almost immediately due to a lack of radiative 
force. Only part of the mass progresses to larger radii, where a similar process repeats itself.  Whenever an element 
moves outward, the radiative luminosity decreases causing elements further out to slow down and/or fall back. If an 
element falls back, the radiative luminosity increases, which causes elements further out to accelerate. Note that the 
motion of mass elements reverses whenever radiative luminosity is particularly low or high.}
\label{fig:supertir}
\end{figure}

\subsection{Exceeding the photon tiring limit}
By increasing the luminosity to ${\Gamma=10}$ and reducing the porosity length we increase the mass loss rate at the 
stellar surface to more than twice the photon tiring limit (Eq.~78 from OGS). The result can be seen in 
Fig.~\ref{fig:supertir}, which shows both the motion of individual mass elements (lines) and radiative luminosity 
(surface contour) as a function of time. Mass is launched from the stellar surface, but most of it quickly falls back 
as the luminosity can not support it against gravity. Only a fraction of the mass that was initially launched manages 
to continue.
At larger radii it encounters previously ejected mass, which is falling down again. 
Each time a mass element falls down, it re-energizes the radiation field (Eq.~\ref{eq:tir}) which allows the radiation 
to drive the mass element above further outward. If a mass element moves outward the radiation field loses energy, 
which causes the mass element above to decelerate or even fall back down.

This process repeats itself several times, creating layers of mass moving back and forth. Only part of the mass 
eventually reaches the point where its velocity is larger than the local escape velocity, allowing it to escape from 
the gravity of the star.
Clearly, the resulting wind has no stable solution. The mass loss rate of this wind is very high, reaching 80\% of the 
photon tiring limit. However, the wind velocity is very low, since most of the energy is being used to lift this large 
amount of mass against the stellar gravity.

\section{Conclusions and plans for the future}
We have shown that continuum driving can cause a star to lose large amounts of mass in a short period, provided that 
the star exceeds the effective Eddington limit. As long as the mass loss rate stays below the photon tiring limit, 
continuum driving can produce a steady wind with a high density and velocity on the order of the escape velocity. 
Once the mass loss rate exceeds the photon tiring limit the wind no longer has a steady solution. Instead the space 
close to the star becomes a tangle of mass elements moving up and down, depending on the amount of radiative energy 
that is available to drive them. The effective mass loss rate (that fully escapes the star) is still very large, but the wind 
velocity becomes quite low, since most of the available energy is used to lift the mass against gravity.
Since continuum driving does not depend on the presence of metals it provides a very effective means to remove mass 
from zero- and low-metallicity stars. Such stars are also thought to be generally more massive than stars at solar 
metallicity (Schaerer \cite{Sc02}, O'shea \& Norman \cite{ON07}), which makes them more likely to exceed the Eddington 
limit.

In the future we intend to include 2D radiative hydrodynamics in our simulations. We also want to compare our results 
with known super-Eddington objects in the nearby universe, such as LBVs and Novae. Hopefully this will allow us to 
create a framework for simulations of the mass loss rate of low metallicity stars.


\begin{theacknowledgments}
A.J.v.M. acknowledges support from NSF grant AST-0507581. We thank A. ud-Doula and R. Townsend for helpful discussions 
and comments.
All simulations were done with the ZEUS 3D hydrodynamics code \cite{sn92}.
\end{theacknowledgments}



\bibliographystyle{aipprocl} 


\end{document}